\documentclass[letter]{aa} 

\usepackage{graphicx}
\usepackage{txfonts}
\begin{document}

   \title{Gamma--ray burst engines may have no memory}


   \author{A.~Baldeschi\thanks{adriano.baldeschi@student.unife.it}
          \and
          C. Guidorzi\thanks{guidorzi@fe.infn.it}
          }

   \institute{Department of Physics and Earth Sciences, University of Ferrara, via Saragat 1,
  I-44122, Ferrara, Italy\\
             }


 
   \abstract
       {A sizeable fraction of gamma--ray burst (GRB) time profiles consist
         of a temporal sequence of pulses. The nature of this stochastic
         process carries information on how GRB inner engines work.
         The so--called interpulse time defines the interval between adjacent
         pulses, excluding the long quiescence periods during which the signal drops
         to the background level. It was found by many authors in the past that
         interpulse times are lognormally distributed, at variance with the
         exponential case that is expected for a memoryless process.}
       {We investigated whether the simple hypothesis of a temporally uncorrelated sequence
         of pulses is really to be rejected, as a lognormal distribution necessarily implies.}
       {We selected and analysed a number of multi--peaked CGRO/BATSE GRBs and
         simulated similar time profiles, with the crucial difference that we assumed
         exponentially distributed interpulse times, as is expected for a memoryless stationary
         Poisson process. We then identified peaks in both data sets using a novel
         peak search algorithm, which is more efficient than others used in the past.}
       {We independently confirmed that the observed interpulse time distribution is
         approximately lognormal. However, we found the same results on the simulated
         profiles, in spite of the {\em intrinsic} exponential distribution.
         Although intrinsic lognormality cannot be ruled out, this shows
         that {\em intrinsic} interpulse time distribution in real data could still be exponential,
         while the observed lognormal could be ascribed to the low efficiency of peak search
         algorithms at short values combined with the limitations of a bin-integrated profile.}
       {Our result suggests that GRB engines may emit pulses after the fashion of nuclear
         radioactive decay, that is, as a memoryless process.}

   \maketitle
%

\section{Introduction}
\label{sec:intro}
Decades after their discovery, the prompt emission of gamma--ray bursts (GRBs) is
still one of the least understood aspects of the GRB phenomenon. While it is
established today that most (if not all) long-duration GRBs are connected with the final
collapse of some kind of massive stars, the nature of the inner engine and the
process(es) through which $\gamma$--rays are released are unclear
(see \citealt{KumarZhang14rev} for a recent comprehensive review).
GRB prompt emission time-profiles consist of one or several pulses without compelling
evidence for periodic patterns despite occasional claims \citep{Beskin10,Cenko10,DeLuca10}.
The information connected with the time interval between two adjacent pulses, the so--called
interpulse time (IT), and its possible correlations with properties of the same pulses can shed
light on the process at work \citep{RamirezRuiz01a,RamirezRuiz01b}.
Within the context of internal shocks, the IT distribution (hereafter, ITD) is tightly connected
with the emission-time history of the inner engine \citep{Kobayashi97}. Specifically,
it can reveal the nature of the stochastic process ruling the production of elementary
energy releases in the form of pulses, which is likely driven by the way accretion works
in GRB engines (e.g., \citealt{TchekhovskoyGiannios14,Bernardini13}).

The importance of ITDs cannot be overstated, as is the case for other different
astrophysical sources, such as the Sun (e.g., \citealt{Wheatland00,Aschwanden10}),
and outbursting magnetars \citep{Gogus99,Gogus00,Gavril04}.
In the case of GRBs a number of papers have investigated this property
\citep{McBreen94,Norris96}. The different peak detection algorithms
\citep{Li96,NakarPiran02a,DragoPagliara07,Bhat12} or techniques \citep{Quilligan02} yielded as the
main result that the GRB ITD is generally well described by a lognormal
with mean values $\la 1$~s, with evidence for a power--law excess at relatively
long ($\Delta t>5$--$10$~s) ITs. These long, rare ITs during which the GRB signal drops
to background are often referred to as quiescent times (QTs) and are interpreted
as caused by something different from what rules the shorter and more frequent ITs
(e.g., \citealt{Drago08,TchekhovskoyGiannios14}).
The identification of QTs in individual time profiles was made by different authors
in different ways through different operative, loose definitions, similarly to what
was done for emission precursors (e.g., \citealt{Burlon08}), even though a recent
time--frequency algorithm adapted from gravitational wave data analysis seems to
provide an interestingly less subjective alternative \citep{Charisi14}.

So far, most efforts were focused on identifying and interpreting QTs as opposed to ITs, especially when extremely long-duration ($\sim10^2$--$10^3$~s) GRBs
with comparably long QTs are occasionally observed.
This has led part of the GRB community to think in terms of 
a class of their own, known as ultra--long GRBs (\citealt{Gendre13,Levan14,Evans14};
see however \citealt{Virgili13,Zhang14c}).
However, the other side of the ITD, the short IT tail, did not raise much
speculation, although the nature of lognormal, that is, the departure from a pure
exponential distribution, would imply some kind of memory in the
GRB engine. In other words, a pure memoryless engine, when one neglects QTs, gives
an exponential ITD that deviates from lognormality. This means that one observes fewer
short ITs than expected for a memoryless engine.
Before trying to interpret the physical as well as statistical origin of the
lognormal ITD \citep{Ioka02}, it is therefore crucial to establish to what extent
the dearth of short interpulse-times is due to the low efficiency of algorithm(s).
In other words, it is key for understanding whether the lognormal distribution,
particularly at short values, is {\em \textup{intrinsic}}.

Our aim is to test whether the pure memoryless case is to be rejected when the
effects of the peak detection algorithms are carefully taken into account by means
of synthetic light curves. More specifically, we simulate time profiles assuming
exponentially distributed ITs that otherwise resemble real GRB profiles as
closely as possible, and compare the corresponding ITDs with that of real data.
In Sect.~\ref{sec:data} we describe the real data sample selection and how
simulations were carried out. Results and implications are then reported
and discussed in Sects.~\ref{sec:res} and \ref{sec:conc}.

\section{Data analysis}
\label{sec:data}

\subsection{Real data selection}
\label{sec:data_sel}
We started from the light curves of the BATSE catalogue made available by the team
as concatenated burst data with 64 ms temporal resolution in four energy
channels.\footnote{ftp://cossc.gsfc.nasa.gov/compton/data/batse/ascii\_data/64ms/}
We took the summed light curve of the four energy channels and preliminarily
subtracted the background by fitting it with polynomials of up to fourth degree,
as suggested by the BATSE team (e.g., \citealt{Guidorzi05c}),
and then applied {\sc mepsa} \citep{MEPSA,Guidorzi14b}, a novel peak search algorithm.
By selecting the GRBs with at least 20 peaks each we formed a final sample of
85 GRB light curves, which we hereafter refer to as the real sample.
The choice of $20$ peaks is somewhat arbitrary, and it was aimed at ensuring
statistical significance. 
The choice of the BATSE catalogue was driven by the unrivalled wealth of
multi-peaked GRBs. In principle, if there were many pulses
narrower than 64~ms, the peak search would not be optimally set up.
However, most GRBs have minimum variability timescales above
tens of milliseconds \citep{Golkhou14}, and autocorrelation \citep{Fenimore95}
and pulse--fitting \citep{Norris96} studies find typical pulse widths of
$\ga0.2$~s in the BATSE energy ranges.

That we used {\sc mepsa} instead of other algorithms,
such as the popular one proposed by \citet{Li96} (LF), was motivated by its much
lower false-positive probability (FP; 1--2$\times10^{-5}$ to be compared with
3--5$\times10^{-3}$~FP~bin$^{-1}$), particularly when the signal drops to background,
and by its capability to detect slowly varying, dim peaks.
Essentially, {\sc mepsa} simultaneously compares various moving intervals (with different
lengths) with adjacent bins against a number of thresholds (in units of statistical noise)
that must be fulfilled simultaneously to trigger at least one out of 39 different criteria.
This search was carried out over increasingly longer timescales \citep{Guidorzi14b}.
The basic principle is very similar to that of LF, but, in addition, {\sc mepsa} has a lower
FP rate because multiple conditions are to be fulfilled, while its sensitivity to dim and long-lasting
peaks is ensured by its multiple--timescale monitoring.

\subsection{Simulated data set}
\label{sec:simul}
The light curves were simulated assuming exponentially distributed interpulse times.
To this aim, much effort was put so as to ensure that the simulated time-profiles were as alike to
observed GRB profiles as possible.
We assumed the shape by \citet{Norris96} for a single shot, which well reproduces a so--called
fast-rise exponential decay (FRED),
\begin{eqnarray}
f(t) & = &  A\, \exp{[-(|t-t_{\rm p}|/\tau_r)^\nu]}  \quad , (t<t_{\rm p})\nonumber\\
     & = &  A\, \exp{[-(|t-t_{\rm p}|/\tau_d)^\nu]}  \quad , (t>t_{\rm p})\;,
\label{eq:shot}
\end{eqnarray}
where $A$ is the peak counts, $t_{\rm p}$ is the peak time, $\tau_{r,d}$ are the rise
and decay times, and $\nu$ is the so-called peakedness, which determines the pulse sharpness.
Except for $t_{\rm p}$, these parameters were found to be approximately lognormally
distributed \citep{Norris96,Quilligan02},
so we assumed corresponding distributions to reproduce the variety of observed time profiles.
We verified that the uncertainties on these distribution parameters had a negligible
impact on the results.
We generated random light curves for each GRB of the real sample to mimic the average temporal features of the real curve as
closely as possible.

The more two adjacent pulses overlap, the more difficult for any peak search algorithm
to identify them as separate pulses.
The so-called separability, which describes how easily a given pulse can be separated from the adjacent
ones, depends on both the pulse duration and the ITs next to it, and affects
the chance {\sc mepsa} has to identify it. In this context, using algorithms such
as LF instead of {\sc mepsa} exacerbates the situation \citep{Guidorzi14b}.
In addition, the signal--to--noise ratio of the pulse itself affects its detectability.
We therefore ensured that both properties, pulse duration and peak counts, were as
similar as possible to what is observed in the corresponding real GRB curve.
We summarise in the following steps the procedure we adopted:
\begin{enumerate}
\item Simulate an exponential ITD whose expected value equals the mean value of
the real ITD of a given profile from the real sample. The mean value is computed
by excluding possible QTs from the light curve. In Sect.~\ref{sec:QT} we describe how
we defined QTs.
The intrinsic expected value of the real curve is probably shorter (by about a factor $F$)
than the observed value, because very close peaks cannot be resolved.
This turns into a loss of true, short ITs.
Starting from 1, we determined $F$ by assuming progressively decreasing values, until
the overall real and simulated ITDs (as obtained by applying {\sc mepsa}) were compatible
according to a Kolmogorov--Smirnov (KS) test.
$F$ was found only once for all GRBs together and was $0.67$.
\item Simulate lognormally distributed peak counts whose expected value and
variance were preliminarily estimated from the data of the corresponding real light curve. 
\item Simulate lognormally distributed values for $\tau_r$ and $\tau_d$, with
expected values equal to the corresponding observed real values.
The FWHM of one pulse is  $(\tau_r+\tau_d)(\ln{2})^{1/\nu}$.
\item Finally, we added the background observed in the real curve and added
Poisson statistical noise.
\end{enumerate}
The peakedness is not a crucial parameter to our aim, so we fixed it to $\nu=1.5$, as
is typically observed in real data \citep{Norris96}.
The corresponding typical FWHM is therefore $0.78\,(\tau_r+\tau_d)$.

\subsection{Quiescent times}
\label{sec:QT}
For both the real and the simulated ITDs we preliminarily removed the so--called
quiescent times, that is, ITs during which the GRB signal drops to
the background level for a relatively long time
\citep{NakarPiran02a,Quilligan02,DragoPagliara07}.
There is no rigorous definition of QT, since the evidence for them as a 
separate class comes from a power--law tail in the ITD in excess of the
lognormal distribution tail \citep{NakarPiran02a,Quilligan02,DragoPagliara07}.

Hence we adopted the following operative procedure to identify QTs:
Let $t_{{\rm p},i}$ and $t_{{\rm p},i+1}$ be the times of two adjacent peaks.
The interpulse time $\Delta t_i=t_{{\rm p},i+1}-t_{{\rm p},i}$ is said to be
a QT when it fulfils the following requirements:
\begin{enumerate}
\item $\Delta t_i \ge 5$~s;
\item there exists at least one time bin $t_k$ ($t_{{\rm p},i}<t_k<t_{{\rm p},i+1}$) such that
  $c_k\le 3\,\sigma_k$, that is, the signal drops to background within $3\,\sigma$
  ($\sigma_k$ is the error on the background--subtracted counts $c_k$);
\item let $t_{i,1}={\rm min}(t_k)$ and $t_{i,2}={\rm max}(t_k)$, calculated over the
$t_k$ that fulfil 2.;
  let $c_{{\rm M},i}={\rm median}(c_j)$ and $\sigma_{{\rm M},i}={\rm median}(\sigma_j)$
  ($t_{i,1}\le t_j\le t_{i,2}$). The condition to be fulfilled is $c_{{\rm M},i}\le \sigma_{{\rm M},i}$.
\end{enumerate}
While condition 1 demands that the QT candidate is long enough, conditions 2 and 3 ensure
that the median signal is compatible with background.
The lowest value of 5~s was estimated from the time at which the power--law excess
becomes visible with respect to the lognormal tail in the ITD (see also
\citealt{Quilligan02,DragoPagliara07}). 

\section{Results}
\label{sec:res}
%
\begin{figure}
\includegraphics[scale=0.75]{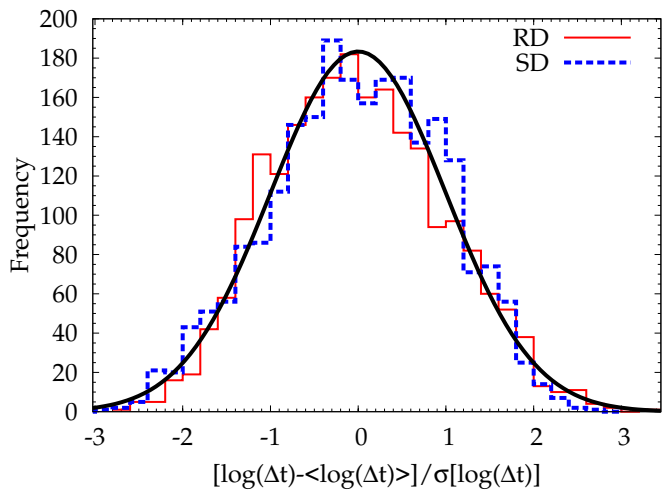}
\caption{Rescaled RD (thin solid) and SD (dashed) ITDs, including 2220 and
2298 interpulse times, respectively. The thick solid line is $N(0,1)$ and is
normalised to the SD sample.}
\label{fig:RITD}
\end{figure} 
Following the procedure described in Sect.~\ref{sec:data}, we obtained three different ITDs for each event of our
real sample: that on the real curve (real detected, RD),
the true one based on the peak times as they were
generated in the simulated curve (simulated true, ST), and the one obtained by applying
{\sc mepsa} to the simulated curve (simulated detected, SD).
We detected 2220 ITs in the real sample in total. In the ST sample we generated 3548
ITs, 2298 of which were measured with {\sc mepsa}. Thus, RD, ST, and SD ITDs
collect 2220, 3548, and 2298 ITs, respectively.
For each simulated curve the peak search was carried out over a time interval with the
same duration as that of the interval considered for the corresponding real curve.
The different size of the RD and SD samples is 78, which is compatible with expectations
from Poisson statistics, $\sqrt{2220+2298}\sim67$.

Since each real GRB has its own average peak rate, we had to rescale based on each individual mean rate before
merging the corresponding ITD. Following \citet{Li96}, we
rescaled the RD and SD ITDs of each GRB by subtracting the logarithmic mean and rescaling
by the sample variance.
This way, each individual ITD has null mean and unitary variance.
We finally merged the entire sample into RD and SD total ITDs for all
GRBs together, which we refer to as rescaled ITDs; they are displayed in Fig.~\ref{fig:RITD}.
The rescaled (non--rescaled) RD and SD are indistinguishable according to a
KS test, whose p--value is $9.2$\% ($30$\%), and both appear to be lognormally distributed.
Given the large number of ITs in each distribution, the KS test is correspondingly sensitive.
Therefore, one cannot reject the hypothesis that the corresponding exponential ST ITD may
also well be the {\em \textup{intrinsic}} distribution of real data.
Figure\ref{fig:RITD} shows that the rescaled SD and RD ITDs look broadly compatible
with a lognormal, although a rigorous $\chi^2$ test for lognormality yields
$<1$\% p--values in both cases.

\begin{figure}
\includegraphics[scale=0.75]{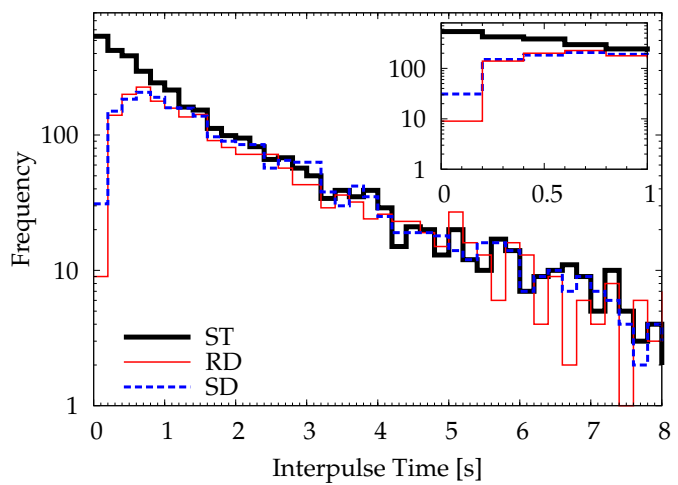}
\caption{ITDs of real detected (thin solid), simulated true (thick solid),
and simulated detected (dashed). A close--up in the short interpulse--time region is shown
in the inset.} 
\label{fig:ITD}
\end{figure} 
The effect of the peak detection algorithm efficiency on the observed distribution is shown
by the comparison between ST and SD in Fig.~\ref{fig:ITD}.
At $\Delta t<0.8$~s (that is, 12 times the temporal bin size of 64~ms) the SD significantly
deviates from the ST ITD: the dearth of short ITs is due to
the low efficiency of {\sc mepsa} when nearby pulses become hardly separable \citep{Guidorzi14b}.

We also explored the compatibility of an intrinsic lognormal ITD with real data.
We carried out the same simulations and found that the best match between rescaled
(non--rescaled) SD and RD yielded a KS p--value of $32$\% ($4.2$\%).

Taken at face value, these results indicate that the lognormal ITD observed
in real data, in particular the dearth of short values compared with what is expected for
a (memoryless) stationary Poisson process, might be intrinsic, but it might also be
an artefact of the low efficiency of the peak detection algorithms, even at relatively long
ITs compared with the binning time of the profile ($\Delta t\la 10\,\times$64~ms).
Thus, lognormality is not necessarily an intrinsic property of the real ITD.

\begin{figure}
\includegraphics[scale=0.70]{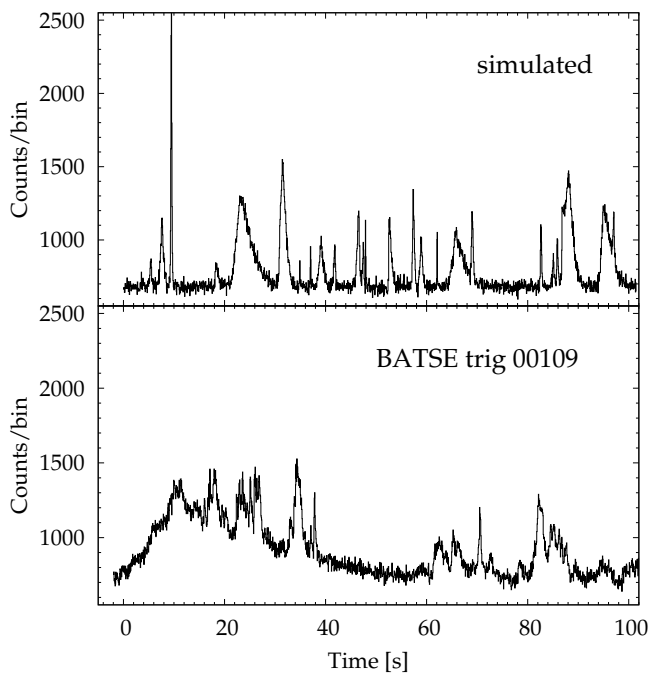}
\caption{Example of a real curve ({\em bottom}) and of its corresponding
simulated one ({\em top}) assuming an exponential ITD.}
\label{fig:curves}
\end{figure} 
Figure 3 illustrates an example of real and corresponding simulated curves.
In the real curve there is a slowly varying component superposed on peaks that have no counterpart in the simulations. This kind of low pedestal is strongest in the
softest energy channels and almost absent from the hardest channels, as was first
pointed out in the hard X--ray (2-28 keV) prompt emission of {\em BeppoSAX} GRBs \citep{Vetere06}.
For our purposes, its absence in simulated curves is not a problem since ITs are given by
the spikier activity in the real and simulated data.

\section{Discussion and conclusions}
\label{sec:conc}
Using a novel peak search algorithm, we confirmed previous results that ITDs in multi--peaked GRBs are approximately lognormal when the long--tail of QTs is not considered.
Surprisingly, it had not been investigated before whether the lack
of short ITs, even at $\Delta t$ a factor of a few longer than the temporal binning
size, might entirely be an artefact of the low efficiency of peak searches.
We amended this lack through simulated profiles that share common properties
with the observed real profiles in terms of peak counts, pulse durations, and measured
ITs, as proved by KS tests. What is more, we proved that assuming an intrinsic
memoryless process in which GRB engines emit pulses according to a stationary
Poisson process on timescales $\Delta t<5$--$10$~s (that is, neglecting QTs),
the true exponential ITD still cannot be recovered because of the
low efficiency of peak detection algorithms in the short IT tail.
In particular, we were able to reproduce a {\em \textup{measured}} ITD that is compatible
with the real one and approaches lognormality, in spite of its intrinsic
exponential nature.

While lognormality is reminiscent of central limit theorem and probably is the
result of the multiplicative combination of independent variables \citep{Ioka02}, our results show that the simplest hypothesis one can make about the way GRB
engines work, that is, an uncorrelated process on relatively short time intervals,
cannot be rejected.

The immediate implication is that GRB engines possibly do not retain any
memory after they have just emitted a pulse and do not, {\em \textup{on average}},
wait for some more time before emitting the next one.
We therefore conclude that the lognormality of the observed ITDs in real GRBs
may not be intrinsic, since it can entirely be explained as an artefact of
peak-finding algorithms.
In terms of accretion process, our results suggest that the individual accretion
episodes that shape the GRB prompt emission profiles in terms of pulses may
on average be uncorrelated, as long as long quiescent periods are not considered.
This can help constrain the way magnetic flux repeatedly induces and halts
accretion, if it is verified \citep{Bernardini13,TchekhovskoyGiannios14}.
Another possibility is that current--driven instabilities in strongly magnetised
jets can lead to non-axisymmetric jet structures which, combined with beaming
effects and even small changes in the Lorentz factor, are responsible for the
observed short timescale variability \citep{Levinson13}.
Alternatively, short timescale variability could result from magnetic
reconnection episodes triggered by collisions between magnetised shells
\citep{ICMART}, which do not necessarily require temporal correlation,
in the same way that internal shocks do not, either.

\begin{acknowledgements}
  We are grateful to the referee for very useful comments that helped improve the paper.
  We also thank S.~Dichiara, F.~Frontera, P.~Rosati, L.~Amati, A.~Drago
  for useful discussions. PRIN MIUR project on ``Gamma Ray Bursts: from
  progenitors to physics of the prompt emission process'', P.~I. F. Frontera
  (Prot. 2009 ERC3HT) is acknowledged.
\end{acknowledgements}


\end{document}